\documentclass[unsortedaddress,twocolumn,pre,amsmath,amssymb]{revtex4}
\usepackage[dvips]{graphicx}% Include figure files
\usepackage{dcolumn}% Align table columns on decimal point
\usepackage{bm}
\begin{document}
\title{A stochastic optimal velocity model and its long-lived metastability}
% Force line breaks with \\
\author{Masahiro Kanai}%
% \email{kanai@ms.u-tokyo.ac.jp}
 \affiliation{%
Graduate School of Mathematical Sciences, 
University of Tokyo, 3-8-1 Komaba, Tokyo 153-8914, Japan.
}%
\author{Katsuhiro Nishinari}
% \email{knishi@rins.ryukoku.ac.jp}
\affiliation{%
Department of Aeronautics and Astronautics, Faculty of Engineering, University of Tokyo, 7-3-1 Hongo, Tokyo 113-8656, Japan.
}%
\author{Tetsuji Tokihiro}%
% \email{toki@poisson.ms.u-tokyo.ac.jp}
 \affiliation{%
Graduate School of Mathematical Sciences, University of 
Tokyo, 3-8-1 Komaba, Tokyo 153-8914, Japan.
}%
\date{\today}% It is always \today, today,
\begin{abstract}
In this paper, we propose
 a stochastic cellular automaton model of traffic flow
 extending two exactly solvable stochastic models, i.e.,
 the asymmetric simple exclusion process and the zero range process.
Moreover it is regarded as
 a stochastic extension of the optimal velocity model.
In the fundamental diagram (flux-density diagram), our model
 exhibits several regions of density where more than one stable state
 coexists at the same density
 in spite of the stochastic nature of its dynamical rule.
Moreover, we observe that two long-lived metastable states appear
 for a transitional period, and that the dynamical phase transition
 from a metastable state to another metastable/stable state occurs
 sharply and spontaneously.
\end{abstract}
%%%%%%%%%%%%%%%%%%%%%%%%%%%%%%%%%%%%%%%%%%%%%%%%%%%%%%%%%%%%%%%%%
\maketitle
%%%%%%%%%%%%%%%%%%%%%%%%%%%%%%%%%%%%%%%%%%%%%%%%%%%%
Traffic dynamics has been attracting much attention
 from physicists, engineers and mathematicians
 as a typical example of non-equilibrium statistical mechanics
 of self-driven many particle systems
 for the last decade \cite{helbing,css,nagatani}.
Statistical properties of traffic phenomena
 are studied empirically by using the {\it fundamental diagram},
 which displays the relation of the flux
 (the average velocity of vehicles multiplied by the density of them)
 to the density.
We have been found 
 an emergence of more than one different flux at the same density
in the transit region from free to congested phase
 is almost universal in real traffic \cite{NH}.
Recent experimental studies also show that the phase transition occurs
 discontinuously against the density
 and structurally complex states appear around
 the critical density \cite{Kerner,THH,NFS}.

%%%%%%%%%
While traffic models with many parameters and complex rules
 may reproduce empirical data, a strong mathematical support, if any,
 allows a direct connection between microscopic modelling
 and the universal feature extracted from various traffic flows.
In this paper, we therefore propose a cellular automaton (CA) model
 extending two significant stochastic processes, i.e.,
 the asymmetric simple exclusion process (ASEP)
 and the zero range process (ZRP) as described later.
Moreover, we find that our model is supported
 by a successful traffic model,
 {\it the optimal velocity (OV) model} \cite{Bando,Bando2}.
The OV model, which is a continuous and deterministic model,
 is expressed by coupled differential equations;
$d^2 x_i/dt^2=a[V(x_{i+1}-x_i)-d x_i/dt]$,
%\begin{equation}
% \frac{d^2 x_i}{dt^2}=a\Bigl[V(x_{i+1}-x_i)-\frac{d x_i}{dt}\Bigr],
%\label{OV}
%\end{equation}
where $x_i=x_i(t)$ is the position
 of the $i$-th vehicle at time $t$
 and the function $V$ is called the optimal velocity function,
 which gives the optimal speed of a vehicle
 according to its headway $x_{i+1}-x_i$
 (the $i$-th vehicle follows the $(i+1)$-th in the same lane).
The intrinsic parameter $a$ indicates the driver's sensitivity
 to traffic situations and governs the stability of flow.
Our stochatic CA model is, however, different from
 a noisy OV model \cite{HS} as described below.%in the following.

First of all, we explain the general framework
 of our stochastic CA model for one-lane traffic.
$N$ vehicles are moving on a single-lane road
 which is divided into a one-dimensional array of $L$ sites.
Each site contains one vehicle at most,
 and collision and overtaking are thus prohibited
 (the so-called {\it hard-core exclusion rule}).
In this paper, parallel updating is adopted,
 i.e., all the vehicles attempt to move at each step.
We introduce a probability distribution function $w^t_i(m)$
 which gives the probability,
 or the driver's {\it intention}, of the $i$-th vehicle
 hopping $m~(m=0,1,2,\ldots)$ sites ahead at time $t$.
Then assuming that the next intention $w^{t+1}_i(m)$ is determined
 by a function $f_i$ depending on $w^t_i(0),w^t_i(1),\ldots$
 and the positions $x^t_1,x^t_2,\ldots,x^t_N$,
 the configuration of vehicles is recursively updated
 according to the following procedure:
\begin{itemize}
\item For each vehicle, calculate the next intention to hop $m$ sites
 with the configuration $x^t_1,x^t_2,\ldots,x^t_N$
 and the intention $w^t_i(0),w^t_i(1),\ldots$;
\begin{equation}
 w^{t+1}_i(m)=f_i\bigl(w^t_i(0),w^t_i(1),\ldots;x^t_1,\ldots,x^t_N;m\bigr).\label{gen}
\end{equation}
\item Determine the number of sites ${\sf V}^{t+1}_i$
 at which a vehicle moves (i.e. the velocity)
 probabilistically according to the intention $w^{t+1}_i$.
In other words,
 the probability of ${\sf V}_i^{t+1}=m$ is equal to $w^{t+1}_i(m)$.
\item Each vehicle moves avoiding a collision;
\begin{equation}
\begin{gathered}
 x^{t+1}_i=x^t_i+\min(\Delta x^t_i,\,{\sf V}^{t+1}_i),\label{genX}
\end{gathered}
\end{equation}
 where $\Delta x^t_i=x^t_{i+1}-x^t_i-1$ defines the headway,
 and the vehicles thus move
 at either ${\sf V}^{t+1}_i$ or $\Delta x^t_i$ sites.
If ${\sf V}^{t+1}_i>\Delta x^t_i$,
 a vehicle must stop at the cell $x^t_{i+1}-1$.
\end{itemize}

%%%%%%%
In what follows, we assume
 that the maximum allowed velocity is equal to 1, i.e.,
 if $m\geqslant2$, $w^t_i(m)=0$.
Then, putting $v^t_i\equiv w^t_i(1)$
 (accordingly $w^t_i(0)=1-v^t_i$),
 we propose a special type of $f_i$ in (\ref{gen}) as
\begin{equation}
 v^{t+1}_i=(1-a_i)v^t_i+a_iV_i(\Delta x^t_i)\qquad 
(\forall t\ge0,~\forall i),\label{SOV}
\end{equation}
 where $a_i~(0\le a_i\le 1)$ is a parameter
 and the function $V_i$ is restricted to values in the interval
 $[0,1]$ so that $v^t_i$ also should be within $[0,1]$.
The intrinsic parameter $a_i$, a weighting factor
 of the optimal velocity $V_i(\Delta x^t_i)$
 to the intention $w^{t+1}_i$,
 corresponds to the driver's sensitivity to a traffic condition.
As long as the vehicles move separately,
 we can also rewrite (\ref{genX}) simply as $x^{t+1}_i=x^t_i+1$
 with probability $v^{t+1}_i$.
(Note that the original OV model does not support
 the hard-core exclusion \cite{Bando,Bando2}.)
Therefore, $v^t_i$ can be regarded as the average velocity,
 i.e., $\langle x^{t+1}_i\rangle=\langle x^{t}_i\rangle+v^{t+1}_i$
 in the sense of expectation values.
We call the model expressed by (\ref{SOV})
 {\it the stochastic optimal velocity (SOV) model}
 because of its formal similarity
 to a discrete version of the OV model (or a coupled map lattice\cite{YK})
\begin{align}
 x^{t+1}_i=&x^t_i+ v^{t+1}_i\Delta t,\label{dOVx}\\
 v^{t+1}_i=&(1-a\Delta t)v^{t}_i+(a\Delta t)V(\Delta x^{t}_i),
\label{dOVv}
\end{align}
 where $\Delta t$ is a time interval and the OV model is recovered
 in the limit $\Delta t\rightarrow 0$.
In this special case that the maximum allowed velocity equals to 1,
 we thus have an obvious correspondence of our stochastic CA model
 to an existent traffic model.
As a matter of convenience, we set $a_i=a$ and $V_i=V~(\forall i)$ hereafter.

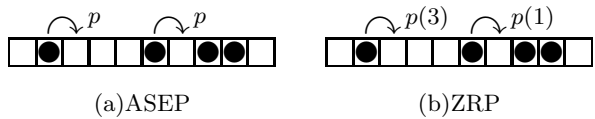
\begin{figure}[t]
\begin{center}
\begin{picture}(220,30)(0,-6)
\multiput(0,0)(10,0){10}{\framebox(10,10)}
\put(15,5){\circle*{8}}
\put(15,10){\makebox(12,10){\Large\bf$\curvearrowright$}}
\put(28,12){\makebox(8,10){$p$}}
\put(55,5){\circle*{8}}
\put(55,10){\makebox(12,10){\Large\bf$\curvearrowright$}}
\put(68,12){\makebox(8,10){$p$}}
\put(75,5){\circle*{8}}
\put(85,5){\circle*{8}}
\put(30,-20){\makebox(40,10){(a)ASEP}}
\multiput(120,0)(10,0){10}{\framebox(10,10)}
\put(135,5){\circle*{8}}
\put(135,10){\makebox(12,10){\Large\bf$\curvearrowright$}}
\put(148,12){\makebox(20,12){$p(3)$}}
\put(175,5){\circle*{8}}
\put(175,10){\makebox(12,10){\Large\bf$\curvearrowright$}}
\put(188,12){\makebox(20,12){$p(1)$}}
\put(195,5){\circle*{8}}
\put(205,5){\circle*{8}}
\put(150,-20){\makebox(40,10){(b)ZRP}}
\end{picture}
\end{center}
\caption{Schematic view of the tagged-particle model for ASEP(a) and ZRP(b).
The hopping probability of a particle depends on the gap size in front
of it in ZRP, while it is always constant in ASEP. In both cases, 
hopping to an occupied cell is prohibited.}
\label{asepzrp}
\end{figure}

%%%%%%%%%
{From} the viewpoint of mathematical interest,
 the SOV model includes two significant stochastic models.
When $a=0$, (\ref{SOV}) becomes $v^{t+1}_i=v^t_i$, i.e., 
 the model reduces to ASEP \cite{Derrida1,Derrida2,Rajewsky,Shutz}
 with a constant hopping probability $p\equiv v^0_i$.
When $a=1$, (\ref{SOV}) becomes $v^{t+1}_i=V(\Delta x^t_i)$,
 i.e., the model reduces to ZRP \cite{Spitzer}
 considering the headways $\{\Delta x^t_i\}$
 as the stochastic variables of ZRP.
In ZRP, the hopping probability of a vehicle is determined
 exclusively by its present headway.
Fig.\ref{asepzrp} illustrates the two stochastic models schematically.
%%% Fig. 1 %%%
%%%%%%%%%%%%%
ASEP and ZRP are both known to be {\it exactly solvable}
 in the sense that the probability distribution of the configuration
 of vehicles in the stationary state
 can be exactly calculated \cite{Evans1,Evans2},
 and thus our model admits an exact calculation
 of the fundamental diagram in the special cases.

%%%%%%%%%%%%%%%%%
%%%%% FIG. 2 %%%%%
\begin{figure}[t]
\includegraphics[scale=0.45]{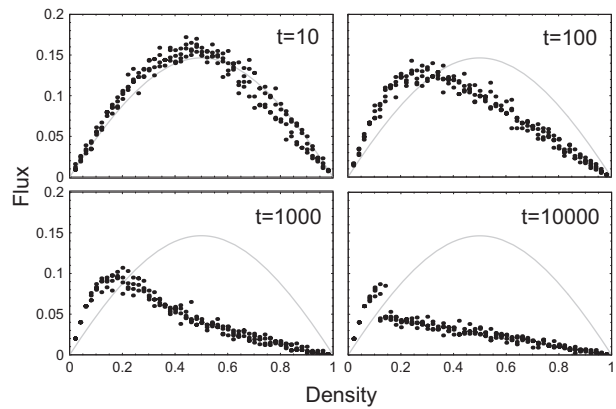}
\caption{The fundamental diagram of the SOV model
 with the OV function (\ref{ov}) (c=$1.5$ and $a=0.01$) plotted
 at each time stage $t$,
 starting from uniform/random states
 with $p\,(\equiv v^0_i)=0.5$,
 including the exact curve(gray) of ASEP %with probability $p=0.5$
 for comparison \cite{Derrida2}.
The system size is $L=1000$.
% and the number of samples is 4 at each density.
}
\label{FD2}
\end{figure}
%%%%%%%%%%%%%%%%
In order to investigate a phenomenological feature, % of the SOV model,
 we take a realistic form of the OV function as
\begin{equation}
\begin{gathered}
V(x)=\frac{\tanh(x-c)+\tanh c}{1+\tanh c},\label{ov}
\end{gathered}
\end{equation}
which was investigated in \cite{Bando}.
We find that the fundamental diagrams simulated
 with (\ref{ov}) have a quantitative agreement
 with the exact calculation of ZRP ($a=1$) up to $a\sim0.6$.
%In contrast, the situation drastically changes
% in the case $0<a<0.1$.
%Although the SOV model coincides with ASEP at $a=0$,
% the fundamental diagram of the SOV model does not come closer
% to that of ASEP as $a$ approaches to 0.
In contrast, the fundamental diagram of the SOV model does not
 come closer to that of ASEP as $a$ approaches to 0
 although the SOV model coincides with ASEP at $a=0$.
Fig. \ref{FD2} shows that
 a curve similar to the diagram of ASEP
 appears only for the first few steps ($t=10$)
 and then changes the shape rapidly ($t=100,\,1000$).
Surprisingly, when the diagram becomes stationary,
 it allows a discontinuous point
 and two overlapping stable states
 around the density $\rho\sim0.14$ ($t=10000$).%(Fig. \ref{FD2} ($t=10000$)).

\begin{figure}[t]
\includegraphics[scale=0.7]{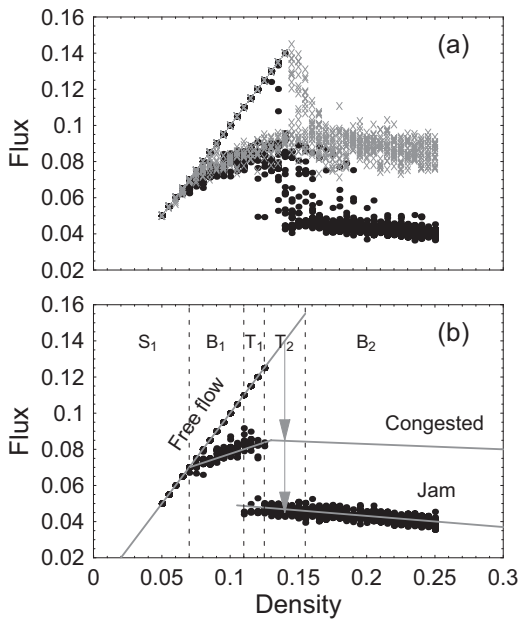}
\caption{
(a)The expanded fundamental diagram of the SOV model with $a=0.01$
 at $t=1000$ (gray) and $t=5000$ (black)
 starting from two typical states;
 the uniform state with equal spacing of vehicles
 and $p\,(\equiv v^0_i)=1$,
 and the random state with random spacing and $p=1$.
We observe three distinct branches, which we call
the free-flow, congested, and jam branch.
They survive even in
the stationary state, which is plotted at $t=50000$ in (b).
(b)The stationary states (black)
 and the averaged three branches (gray lines) are plotted at $t=50000$.
The vertical dotted lines distinguish
 the regions of density from $S_1$ to $B_2$.
The arrows in $T_2$ indicate the trace
of a metastable free-flow state decaying to the lower 
branches. (see also Fig. \ref{spatemp}).
}\label{FDex}
\end{figure}

%%%%%%%%
Let us study the discontinuity of the flux in detail.
Fig. \ref{FDex} shows the fundamental diagram expanded
 around the discontinuous point.
%%%% Fig. 3 %%%%
%%%%%%%%
We have three distinct branches and then call them as follows:
 {\it free-flow phase} (vehicles move without interactions),
 {\it congested phase} (a mixture of small clusters
 and free vehicles), and {\it jam phase} 
(one big stable jam transmitting backward).
These branches appear in the fundamental diagram, respectively
 as a segment of line with slope 1(free-flow),
 as a thick curve with a slight positive slope(congested),
 and as a thick line with a negative slope(jam).
Note that the congested and jam lines shows some fluctuations
due to a randomness of the SOV model.
%%%%%%%%%%%%%%%%%%%%%%%%%
Fig. \ref{FDex}(a) shows snapshots of the flux
 at $t=1000$ and $5000$.
There exists midstream flux
 between free-flow and congested phases at $t=1000$,
 and between congested and jam phases at $t=5000$.
This suggests
 that the high-density free-flow states can hold
 until $t=1000$ but not until $t=5000$,
 and that the high-density congested states have already started
 to decay into jam states before $t=5000$. 
We have thereby revealed the existence of two metastable branches
 leading out of the free-flow or congested lines.
%%%%%%%%%%%%%%%%%%%%%%%%%%
Comparing Fig.\ref{FDex}(a) with Fig. \ref{FDex}(b),
 we have six qualitatively distinct regions
 of density(Fig.\ref{FDex}(b));
 {\it free region $S_1$} (including only free-flow phase),
 {\it bistable region $B_1$} (including free-flow and congested phases, which are both stable),
 {\it tristable region $T_1$} (including all the three phases, which are all stable),
 {\it tristable region $T_2$} (including free-flow, congested and jam phases. The former two phases are stable and the last one is metastable),
 {\it bistable region $B_2$} (including congested and jam phases.
 The former is metastable, and the latter is stable),
 and {\it jam region $S_2$} (including only stable jam phase,
 which is not displayed here).
We stress that the tristable region $T_1$
 is a novel and remarkable characteristic of many-particle systems,
 and that the successive phase transitions from a free-flow state
 to a jam state via a congested state occur respectively
 on a short time scale.

%%%%%%%%%%%%
%%%%%% Fig.4 %%%%%%%
\begin{figure}[t]
\includegraphics[scale=0.7]{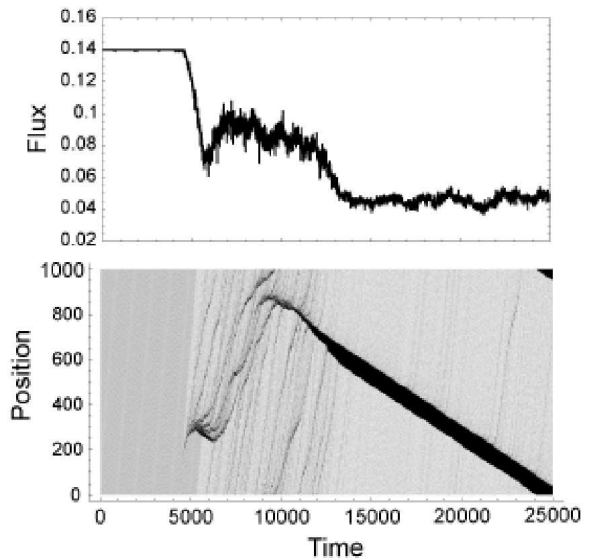}
\caption{The time evolution of flux at the density $\rho=0.14$
 starting from the uniform state.
We observe two plateaus at the flux $Q=0.14$
 with a lifetime $T\simeq 5000$,
 and $Q\simeq0.08$ with $T\simeq 7000$
 before reaching the stationary jam state(upper).
The lower figure shows the corresponding spatio-temporal diagram,
 where vehicles (black dots) move from bottom up.
(Note that the periodic boundary condition is imposed.)}
\label{spatemp}
\end{figure}
\begin{figure}[t]
\includegraphics[scale=0.4]{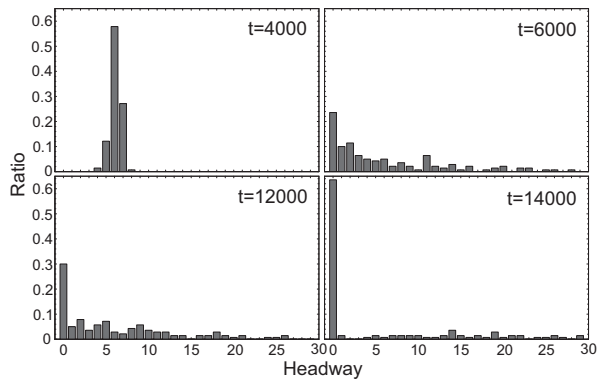}
\caption{The distribution of the headways with which the vehicles move
 at each time stage $t$.
It changes a lot after phase transitions($t\sim5000,12000$).
The traffic states, free-flow($t<5000$), congested($5000<t<12000$),
 and jam($t>12000$) respectively show their specific pictures.
}
\label{bar}
\end{figure}

%%%%%%%%%%%%%%%%%
Let us study the {\it dynamical phase transition} especially
 at the density $\rho=0.14$,
 indicated by successive arrows in Fig. \ref{FDex}(b).
Fig. \ref{spatemp}(upper) shows the flux plotted against time.
It is striking that there appear three plateaus
 which respectively correspond to a free-flow state,
 a congested state, and a jam state,
 and that the flux changes sharply from one plateau to another.
In other words, the metastable states have
 a remarkably long lifetime
 before undergoing a sudden phase transition.
Stochastic models, in general, are not anticipated
 to have such long-lived metastable states
 because stochastic fluctuations break
 a stability of states very soon \cite{NFS}.
Fig. \ref{spatemp}(lower) shows
 the spatio-temporal diagram corresponding to
 the dynamical phase transition.
Starting from a free-flow state, the uniform configuration
 stochastically breaks down at time $t\sim 5000$,
 and then the free-flow state is rapidly replaced
 by a congested state where a lot of clusters are forming
 and dissolving, moving forward and backward.
Fig. \ref{bar} shows the distribution of headways
 at several time stages.
We find that the distribution of headways changes
 significantly after each phase transition.
In particular, the ratio of vehicles with null headway increases.
As for a congested state, it is meaningful to evaluate
 the average size of clusters from a distribution of headways.
If the ratio of the vehicles with null headway is $b_0$,
 the average cluster size $\ell$ can be evaluated as $1/(1-b_0)$.
In the present case, the average size of clusters is
 about $1.2\sim1.4$ during $t=6000\sim12000$.
We also have some remarks, from a microscopic viewpoint,
 on a single cluster:
Since the sensitivity parameter $a$ is set to a small value,
 the intention $v^t_i$ does not change a lot
 before the vehicle catches up with the tail of
 a cluster(i.e. $v^t_i\sim1$).
Accordingly, the aggregation rate $\alpha$ is roughly estimated
 at the density of free region
 behind the cluster; $\alpha\sim(1-b_0)\rho$.
In the present case, the aggregation rate averaged
 over the whole clusters of a congested state is $0.10\sim0.12$.
For the same reason, we can estimate the average velocity
 of the front vehicle of a cluster at $(1-a)^{\ell/\delta}$,
 where $\delta$ denotes the dissolution rate
 and $\ell/\delta$ indicates the duration of capture.
Since $\delta$ is also equivalent to the average velocity
 of the front vehicle, it amounts roughly to $1-a\ell$ after all.
In the present case, the dissolution rate averaged
 over the whole clusters of a congested state is $0.86\sim0.88$.
Then, the average lifetime of the clusters $\ell/(\delta-\alpha)$
 is estimated at $1.54\sim1.89$.
%%%% FIG.5 %%%%
%%%%%%%%%%%%%%%
The above estimations are appropriate
 only when clusters are small and spaced apart.
As many clusters arise everywhere and gather,
 a vehicle out of a cluster tends to be caught
 in another cluster again
 before it recovers its intention at full value.
Consequently, the clusters arising nearby reduce
 their dissolution rates,
 and finally grow into a jam moving steadily backward.
The steady transmitting velocity (the aggregation rate)
 is estimated at $-0.055$, coinciding with Fig. \ref{spatemp}.

%%%%%%% Conclusion %%%%%%%
In this paper, beginning with a general scheme, we have proposed
 a stochastic CA model
 to which we introduce a probability distribution function
 of the vehicle's velocity.
It includes two exactly solvable stochastic processes, %(ASEP and ZRP),
 and it is also regarded
 as a stochastic generalization of the OV model.
Moreover, it exhibits the following features:
In spite of a stochastic model,
 the fundamental diagram shows that there coexist two or three stable phases
 (free-flow, congested, and jam) in a region of density.
As the density increases,
 the free-flow and congested states lose stability
 and change into metastable states
 which can be observed only for a transitional period.
Moreover the dynamical phase transition from a metastable state
 to another metastable/stable state,
 which is triggered by stochastic perturbation,
 occurs sharply and spontaneously.
We consider that the metastable state may be relevant to the transient congested state observed in the upstream of on-ramp \cite{Kerner,MitNak}.
Such a dynamical phase transition has not been observed
 in previous works \cite{Bando,Bando2,HS}
 or among existent many-particle systems.
Further studies, e.g., on another choice of the OV function,
 under open boundary conditions,
 and on the general (i.e. multi-velocity) version will be given
 in subsequent publications \cite{KNT}.
%%%%%%%%%%%%%%%%%%%%%%%%%%%%%%%%
%   References       
%%%%%%%%%%%%%%%%%%%%%%%%%%%%%%%%

%%%%%%%%%%%%%%%%%%%%%%%%%%%%%%%%%%%%%%%%%%%%%%%%%%%%%%%%%%%%%
\end{document}